\documentstyle[preprint,aps,eqsecnum]{revtex}

\def\beq#1{\begin{equation} \label{#1}}
\def\eeq{\end{equation}}
\newcommand{\bea}{\begin{eqnarray}}
\newcommand{\eea}{\end{eqnarray}}

\def\be{\begin{equation}}
\def\ee{\end{equation}}
\def\bd{\begin{displaymath}}
\def\ed{\end{displaymath}}
\def\ba{\begin{array}}
\def\ea{\end{array}}

\begin{document}
{
\tighten
%\preprint {\vbox{
% \hbox{WIS-99/26/July-DPP}
% \hbox{TAUP 2583-99}
% \hbox{hep-ph/9907551}
% \hbox{ANL-HEP-PR-99-70}
%}}

\title {  New Predictions for Multiquark Hadron
Masses}

\author{Harry J. Lipkin\,\thanks{Supported
in part by grant from US-Israel Bi-National Science Foundation
and by the U.S. Department
of Energy, Division of High Energy Physics, Contract W-31-109-ENG-38.}}
\address{ \vbox{\vskip 0.truecm}
  Department of Particle Physics
  Weizmann Institute of Science, Rehovot 76100, Israel \\
\vbox{\vskip 0.truecm}
School of Physics and Astronomy,
Raymond and Beverly Sackler Faculty of Exact Sciences,
Tel Aviv University, Tel Aviv, Israel  \\
\vbox{\vskip 0.truecm}
High Energy Physics Division, Argonne National Laboratory,
Argonne, IL 60439-4815, USA\\
~\\harry.lipkin@weizmann.ac.il
\\~\\
}

\maketitle

\begin{abstract}

The recent reported  charmed-strange resonance at 2.32 GeV/c suggests a
possible multiquark state.  Three types of multiquark bound states are
reviewed. A previous model-independent variational approach  considers a
tetraquark with two heavy antiquarks and two light quarks as a heavy
antidiquark with the color field of a quark bound to the two light quarks with
a wave function like that of a heavy baryon. Results indicate that a
charmed-strange tetraquark $\bar c \bar s u d$ or a  bottom-strange tetraquark
$\bar b \bar s u d$ with this ``diquark-heavy-baryion" wave function is not
bound, in contrast to ``molecular-type" $D-K$ and $B-K$  wave functions.
However, a  charmed-bottom tetraquark $\bar c \bar b u d$ might be bound with a
very narrow weak decay mode. A ``molecular-type" $D-B$ state can have an
interesting $B_c \pi$ decay with a high energy pion,
\end{abstract}

} % end tighten

\section {Three Types of Multiquark states }

The recent observation\cite{babar232} of a charmed-strange state at $2.32$ GeV
that decays into $D_s \pi^o$ suggests a possible four-quark state (tetraquark).
\cite{Jaf77b,Lip77,Isg81,LipVar,Bar94,Ric99,bcl}

Three different mass scales are relevant to the description of multiquark
hadrons, the nuclear-molecular scale, the hyperfine or color-magnetic scale and
the diquark scale.

The nuclear scale is characterized by the deuteron, a bound state of two color
singlet hadrons with a reduced mass of 500 MeV, a binding energy of several
MeV  and a radius of $\approx M_\pi$.  The underlying quark structure of the
hadrons plays no role. The kinetic energy of the state confined to this radius
is
\begin{equation}
T_N =  p^2/M_N \approx M_\pi^2/M_N \approx 20 \rm MeV
\end{equation}

No two-meson bound state containing a pion has been found.  The reduced
mass of any such state is too small to be bound in a radius of $\approx M_\pi$
by a similar interaction;  its kinetic energy would be too high.
\begin{equation}
T_\pi \approx M_\pi^2/M_\pi \approx 140 \, \rm MeV
\end{equation}

The two-kaon system with a reduced mass of 250 MeV seems to be on the
borderline,

\begin{equation}
 T_K=  p^2/M_K \approx M_\pi^2/M_K \approx 40 \, \rm MeV
\end{equation}

Suggestions that the $f_o$ and $a_o$ mesons are deuteron-like $K \bar K$
states or molecules are interesting, but controversial. There is no unambiguous
signature  because $K \bar K$ couples to $\pi-\pi$ and $\eta - \pi$ and both
states  break up strongly.

The $D-K$ system with a kinetic energy
\begin{equation}
T_{DK}=  p^2(M_D+M_K)/2M_DM_K \approx M_\pi^2(M_D+M_K)/2M_DM_K \approx 25\, \rm
MeV
\end{equation}
is therefore an attractive candidate for such a
state\cite{Lip77,Isg81,LipVar,Bar94,Ric99,bcl}. The transition for the $I=0 ~
DK $ state to  $D_S \pi$ is isospin forbidden; thereby suggesting a narrow
width.

The color-magnetic scale is characterized by a mass splitting of the order of
400 MeV; e.g. the $K^* - K$ splitting. Recoupling the colors and spins of a
system of two color-singlet hadrons has been shown to produce a gain in
color-magnetic energy\cite{Jaf77b,Lip77,Isg81}. However,  whether this gain in
potential energy is sufficient to overcome the added kinetic energy required
for a bound state is not clear without a specific model.

The diquark scale arises when two quarks are sufficiently heavy to be bound in
the well of the coulomb-like short-range potential required by QCD.  A heavy
antidiquark in a triplet of color SU(3) has the color  field of a quark and can
be bound to two light quarks with a wave function like that of a heavy baryon.
Since the binding energy of two particles in a coulomb field is proportional to
their reduced mass and all other interactions are mass independent, this
diquark binding must become dominant at sufficiently high quark masses.

\section {The Diquark-Heavy-Baryon Model for Tetraquarks}

We now examine the  diquark-heavy-baryon model for states containing heavy
quarks.  Our ``model-independent" approach assumes that nature has already
solved the problem of a heavy color triplet interacting with two light quarks
and given us the answers; namely the experimental masses of the $\Lambda$,
$\Lambda_c$ and $\Lambda_b$. These answers provided by nature can now be used
without understanding the details of the underlying theoretical QCD model. This
approach was first used by Sakharov and Zeldovich\cite{SakhZel} and has been
successfully extended to heavy flavors\cite{PBIGSKY,arkadlat,Karlip}.

The calculated mass can be interpreted as obtained from a variational principle
with a particular form of trial wave function\cite{LipVar}. This model neglects
the color-magnetic interactions of the heavy quarks, important for the
charmed-strange four-quark system at the colormagnetic
scale\cite{Jaf77b,Lip77,Isg81} and is expected to  overestimate the mass of a
$\bar c \bar s u d$ state. Thus obtaining a model mass value above the relevant
threshold shows only that this type of diquark-heavy-baryon wave function does
not produce a bound state; i.e that the heavy quark masses are not at the
diquark scale. The previous results\cite{Lip77,Isg81} at the colormagnetic or
nuclear-molecular scale should be better.  However, the $bc$ system may already
be sufficiently massive to lead to stable diquarks and the model predictions
for the $\bar c \bar b u d$ state may suggest binding.

We first apply this model to a  $\bar c \bar s u d$  state  with a  light
$ud$ pair seeing the color field of the $\bar c \bar s$ antidiquark like the
field of a heavy quark  in a heavy baryon.  The $\bar c \bar s$
antidiquark differs from the $c \bar s$ in the $D_s$ by having a  $Q Q$
potential which QCD color algebra requires\cite{LipVar} to have half the
strength of the $Q \bar Q$ potential in the $D_s$. The tetraquark mass is
estimated by using the known experimental masses of the heavy baryons and heavy
meson with the same flavors and introducing corrections for the difference
between the heavy meson and the heavy diquark.

\beq{Mtet}
M(\bar c \bar s u d) =  m_c + m_s + m_u + m_d +
\langle H_{udQ} \rangle + \langle H_{ud} \rangle +
\langle T_{cs} \rangle _{cs}+ \langle  V_{cs} \rangle _{cs}
\end{equation}
\beq{Mcs}
M(cs) = m_c + m_s +
\langle T_{cs} \rangle_{cs} + \langle  V_{cs} \rangle_{cs}
\end{equation}
\beq{MDs}
M(D_s) = m_c + m_s +
\langle T_{cs} \rangle_{c \bar s} + \langle  V_{c \bar s} \rangle_{c \bar s}
\end{equation}
\beq{MLam}
M(\Lambda) = m_s + m_u + m_d +
\langle H_{ud} \rangle +
\langle H_{udQ} \rangle
\end{equation}
\beq{MLamc}
M(\Lambda_c) = m_c + m_u + m_d +
\langle H_{ud} \rangle +
\langle H_{udQ} \rangle
\end{equation}
where   $H_{ud}$ and $ H_{udQ}$ respectively denote the Hamiltonians
describing the internal motions of the  $ud$ pair and of  the three-body system
of the $ud$ pair and the antidiquark which behaves like a heavy quark, $T_{cs}$
and $ V_{cs} $ denote the kinetic and potential energy
operators for the internal motion of a $cs$ diquark which is the same as that
for a $\bar c \bar s$ antidiquark. The
expectation values are taken with the ``exact" wave function for the model,
with the subscript $cs$
indicating that it is taken with the wave function of a diquark and not of
the $D_s$.
The kinetic energy operator $T_{cs}$ is the same for the
$cs$ diquark and the $D_s$ but the potential energy operators $V_{cs}$ and
$V_{c \bar s} = 2 V_{cs}$  differ by the QCD factor 2.
This difference between $cs$ diquark and $D_s$ wave
functions is crucial to our analysis.

The quark masses $m_q$ are effective constituent quark  masses and not current
quark masses. We follow the approach begun by  Sakharov and
Zeldovich\cite{SakhZel}   who noted that both the difference $m_s-m_u$ between
the effective masses of strange and nonstrange quarks and their ratio $m_s/m_u$
have the same values  when calculated from baryon masses and meson masses.

\begin{equation}
\langle m_s-m_u \rangle_{Bar}= M_\Lambda-M_N=177\,{\rm MeV}
\end{equation}

\begin{equation}
\langle m_s-m_u \rangle_{Mes} =
{{3(M_{K^{\scriptstyle *}}-M_\rho )
+M_K-M_\pi}\over 4} =180\,{\rm MeV}
\end{equation}

\begin{equation}
\left({{m_s}\over{m_u}}\right)_{Bar} =
{{M_\Delta - M_N}\over{M_{\Sigma^*} - M_\Sigma}} = 1.53
\approx
\left({{m_s}\over{m_u}}\right)_{Mes} =
{{M_\rho - M_\pi}\over{M_{K^*}-M_K}}= 1.61
\end{equation}
where the {\em ``Bar"} and {\em ``Mes"} subscripts denote values obtained
from baryons and mesons, respectively.
Similar results have since been found for hadrons containing heavy quarks along
with many more relations using these same effective quark mass values for
baryon magnetic moments and hadron hyperfine
splittings\cite{PBIGSKY,arkadlat,Karlip}. We therefore assume that the values
of the effective quark masses $m_q$ remain the same for all meson and baryon
states in our analysis.

Substituting eqs.(\ref{Mcs} - \ref{MLamc}) into eq. (\ref{Mtet}) gives
\begin{equation}
M(\bar c \bar s u d)
= (1/2)\cdot[M(D_s) + M(\Lambda) + M(\Lambda_c)] +
\langle \delta H_{cs} \rangle
\end{equation}
where  $\langle \delta H_{cs} \rangle$ expresses the difference between
the $D_s$ and the $\bar c \bar s$ wave functions
\begin{equation}
\langle \delta H_{cs} \rangle  =
\langle T_{cs} \rangle_{cs} + \langle  V_{cs} \rangle_{cs} - (1/2)\cdot
[\langle T_{cs} \rangle_{c \bar s} + \langle  V_{c \bar s} \rangle_{c \bar s}]
\end{equation}
To calculate $\langle \delta H_{cs} \rangle$ we first improve on the treatment
of ref\cite{LipVar} and define the Hamiltonian
\begin{equation}
H(\alpha) = \alpha T_{cs}  +  V_{cs}
 = \alpha T_{cs} +  (1/2)\cdot V_{c \bar s}
\end{equation}
This Hamiltonian $H(\alpha)$ is seen to describe both the $cs$ diquark and the
$D_s$
\begin{equation}
M(cs) = m_c + m_s +  \langle H(\alpha) \rangle_{\alpha = 1}
\end{equation}
\begin{equation}
M(D_s) = m_c + m_s + 2\cdot \langle H(\alpha) \rangle_{\alpha = (1/2)}
\end{equation}
\begin{equation}
\langle \delta H_{cs} \rangle  = \langle H(\alpha) \rangle_{\alpha = 1}
- \langle H(\alpha) \rangle_{\alpha = (1/2)}
\end{equation}
To evaluate $\langle \delta H_{cs} \rangle $ we use  the Feynman-Hellmann theorem and
the virial theorem to obtain,
\begin{equation}
{{d}\over{d\alpha}}\cdot \langle H(\alpha) \rangle
= \left\langle {{dH(\alpha)}\over{d\alpha}} \right\rangle =
\langle  T_{cs} \rangle=
\left\langle {{r}\over{2\alpha}}\cdot{{dV_{cs}}\over{dr}} \right\rangle_\alpha
\end{equation}
\begin{equation}
\langle \delta H_{cs} \rangle  = \int _{(1/2)}^1 d\alpha
\left\langle {{dH(\alpha)}\over{d\alpha}} \right\rangle
= \int _{(1/2)}^1 d\alpha
 \left\langle {{r}\over{2\alpha}}\cdot{{dV_{cs}}\over{dr}} \right\rangle_\alpha
\end{equation}
This expression can be simplified by using the Quigg-Rosner logarithmic
potential\cite{QuiggRos} with its parameter $V_o$ determined by fitting the
charmonium spectrum.
\begin{equation}
V^{QR}_{cs} = (1/2)\cdot V_o \cdot log (r/r_o)
\end{equation}
\begin{equation}
\langle \delta H_{cs} \rangle_{QR}  = {{V_o}\over{4}}\int _{(1/2)}^1 {{d\alpha
}\over{\alpha}} = {{V_o}\over{4}}log\, 2 = 126\,{\rm MeV}
\end{equation}
Substituting experimental values then shows $M(\bar c \bar s u d)$
well above the $DK$ threshold\cite{PDG02}
\begin{equation}
M(\bar c \bar s u d) = (1/2)\cdot[M(D_s) + M(\Lambda) + M(\Lambda_c)] +
\langle \delta H_{cs} \rangle = 2685 + 126 = 2811\,{\rm MeV}
\end{equation}
\begin{equation}
M(\bar c \bar s u d)= 2811\,{\rm MeV} \gg
M(D) + M(K) = 2361\,{\rm MeV}
\end{equation}

In the limit of very high  heavy quark masses this model must give a stable
bound state.  The $cs$ diquark is evidently not heavy enough to produce a bound
diquark-heavy-baryon state.

A similar calculation for $\bar b \bar s u d $ indicates that the
$bs$ diquark is also not heavy enough.
\begin{equation}
M(\bar b \bar s u d) = (1/2)\cdot[M(B_s) + M(\Lambda) + M(\Lambda_b)] +
\langle \delta H_{bs} \rangle = 6180\,{\rm MeV} \gg
%\end{equation}
%\begin{equation}
M(B) + M(K) = 5773\,{\rm MeV}
%\ll  6180\,{\rm MeV}
\end{equation}

However, the $bc$ diquark may be heavy enough to produce a bound four-quark
state.

\begin{equation}
M(\bar c \bar b u d) = (1/2)\cdot[M(B_c) + M(\Lambda_b) + M(\Lambda_c)] +
\langle \delta H_{cs} \rangle = 7280 \pm 200\,{\rm MeV}
\end{equation}
\begin{equation}
M(D) + M(B) = 7146\,{\rm MeV}
\end{equation}

Here the experimental error on the $B_c$ mass is too large to enable any
conclusions to be drawn. But if the bound state exists, it may produce striking
experimental signatures.

A bound $\bar c \bar b u d $, $\bar c \bar b u u $ or $\bar c \bar b d d $
state would decay only weakly. either by b-quark decay into two charmed mesons
(with the same sign of charm, so that there cannot be a J/psi decay mode), or a
c-quark decay into a b meson and a strange meson.  The signature with a vertex
detector will see a secondary vertex with a multiparticle decay and one or two
subsequent heavy quark decays and either one track or no track from the primary
vertex to the secondary.

On the other hand, if the 2.32 GeV state seen by BaBar is really a $DK$ $I=0$
molecule with an isospin violating $D_s-\pi$ decay, the analog for the $bc$
system is a $BD$ molecule with either $I=1$ or $I=0$ and a $B_c-\pi$ decay.
which is isospin conserving for $I=1$ or isospin violating for $I=0$.

Here the masses are very different and give a completely different
signature with a high energy pion. M(B) = 5279 MeV, M(D) = 1867 MeV. This
gives $M(B) + M(D) = 7146$ MeV, while $M(B_c) = 6400 \pm 400$ MeV. So a
molecule just below $BD$ threshold would just rearrange the four quarks into
Bc-pi and fall apart, either with or without isospin violation, giving a
neutral or charged pion having a well defined energy of $750 \pm 400$ MeV
with the precision improved by better measurements.

In any case this is a striking signal which cannot be confused with a $q \bar
q$ state. Experiments can look for a resonance with a pion accompanying any of
$B_c$ states.

It is a pleasure to thank E. L. Berger, T. Barnes, F. E. Close, M.
Karliner, J. Napolitano and V. Papadimitriou  for helpful discussions.

{\tighten

 \end{document}